\RequirePackage{fix-cm}
\documentclass[smallextended]{svjour3}       
\smartqed  
\usepackage{graphicx}
\usepackage{xcolor}
\usepackage{comment}
\begin{document}

\title{Computational model on COVID-19 Pandemic using Probabilistic Cellular Automata 
}
\subtitle{}


\author{       Sayantari Ghosh  \and
        Saumik Bhattacharya$^*$ 
}


\institute{Sayantari Ghosh \at
              Department of Physics, National Institute of Technology Durgapur \\
              \email{sayantari.ghosh@phy.nitdgp.ac.in}           
           \and
           Saumik Bhattacharya ($^*$ corresponding author)\at
              Department of Electronics and Electrical Communication Engineering, Indian Institute of Technology Kharagpur\\
              \email{saumik@ece.iitkgp.ac.in}
}

\date{Received: date / Accepted: date}

\maketitle

\begin{abstract}
Coronavirus disease (COVID-19) which is caused by SARS-COV2 has become a pandemic. This disease is highly infectious and potentially fatal, causing a global public health concern. To contain the spread of COVID-19, governments are adopting nationwide interventions, like lockdown, containment and quarantine, restrictions on travel, cancelling social events and extensive testing. To understand the effects of these measures on the control of the epidemic in a data-driven manner, we propose a probabilistic cellular automata (PCA) based modified SEIQR model. The transitions associated with the model is driven by data available on chronology, symptoms, pathogenesis and transmissivity of the virus. By arguing that the lattice-based model captures the features of the dynamics along with the existing fluctuations, we perform rigorous computational analyses of the model to take into account of the spatial dynamics of social distancing measures imposed on the people. Considering the probabilistic behavioral aspects associated with mitigation strategies, we study the model considering factors like population density and testing efficiency. Using the model, we focus on the variability of epidemic dynamics data for different countries, and point out the reasons behind these contrasting observations. To the best of our knowledge, this is the first attempt to model COVID-19 spread using PCA that gives us both spatial and temporal variations of the infection spread with the insight about the contributions of different infection parameters.
\keywords{Mathematical model of epidemiology \and Lattice epidemic \and Probabilistic cellular automata}
\end{abstract}

\section{Introduction}
\label{intro}
COVID-19 is emerging as one of the biggest pandemics in current times. Affecting 213 countries and territories around the world with a death toll surpassing 400,000 worldwide, COVID-19 is causing a global panic and turmoil \cite{cucinotta2020declares}. SARS-COV2, the causative agent of this disease causes respiratory infection including pneumonia, cold, sneezing and coughing \cite{jin2020epidemiological,pan2020clinical}. It is also found to cause diarrhea, upper respiratory maladies, kidney dysfunction and heart damage in cases of severe infection  \cite{cheng2020kidney,han2020digestive,zheng2020covid}. The disease is highly infectious and it transmits from person-to-person through close contacts, via respiratory droplets produced by infected person while coughing or sneezing \cite{wang2020immediate}. Fomites are also being considered as a significant source of transmission, as the virus has been found to persist on different surfaces up to 72 hours \cite{van2020aerosol}.\\
Mathematical models have long been associated with the study of infectious diseases and diffusion dynamics \cite{bhattacharya2019viral,diekmann2000mathematical}. However, choosing the appropriate model and proper physically relatable parameters have always been a challenge for understanding and predicting the course of a disease. In case of modeling an infectious disease through compartmentalized social structure, the most popular approach is following Kermack and McKendrick \cite{kermack1927contribution} with an ordinary differential-equation (ODE) based SIR (susceptible-infected-recovered) model. The model assumes that when an infectious disease attacks a community, the disease often partitions the population into three subpopulations: individuals who can be infected (\textit{susceptible} people, denoted by $S$) ; people who are already infected (and thus, \textit{infectious}, denoted by $I$); and those who recovered and possess immunity to or got killed by (thus \textit{removed}, denoted by $R$) this disease. Each infected individual transmits the disease with some probability to each susceptible individual they encounter. The infected people recover at a constant rate. In case of COVID-19, some studies have already attempted to gauge the true potential of the disease through SIR model \cite{liu2020reproductive,shim2020transmission}. However,  this simple three compartment model might not be sufficient to appropriately understand the actual nature of this particular infection spread.\\
There are two major divergences of the SIR picture from the case of COVID-19. Firstly, in the case of SARS-COV2, asymptomatic transmission of the virus remained a controversial topic for some time \cite{bai2020presumed,nishiura2020estimation}. Asymptomatic proportion is broadly defined as the fraction of people, who came in contact with the infected, but they do not show any symptoms for some time. For COVID-19, there have been reports of exposure to asymptomatic people resulting into transmission during the incubation period \cite{yu2020familial}. Thus, it can be appreciated that reliable estimate of the asymptomatic proportion can play a notable role to control the intensity and range of the disease. This is critically desired to direct public health policy and social distancing strategies to fight COVID-19. Several studies \cite{bai2020presumed,nishiura2020estimation,rocklov2020covid} reported that the proportion of people, who after remaining asymptomatic for an incubation period have shown prominent symptoms of the infection, lies between 18\% to 41\%. This is a quite significant proportion which cannot be ignored for a highly infectious disease. Thus, while modeling the dynamics, the asymptomatic proportion has to be considered indeed to measure the severity of the problem and better evaluate the transmission potential.\\
Another important factor that cannot be considered in an SIR model structure, is the effect (or lack) of efficient healthcare. The governments of most of the countries are allocating huge funds and manpower for scaling up the hospital facilities to provide nasal oxygen, mechanical ventilation and, for patients with complications, dialysis, as per the clinical guideline suggestions \cite{alhazzani2020surviving,murthy2020care}. However, for several countries, where healthcare systems were already overwhelmed by the demand prior to the pandemic, providing quick and proper testing facility to every infected person is not realistically possible \cite{spinelli2020covid}. Even in countries with outstanding health infrastructure (e.g., Italy, Spain etc.) at the peak of the pandemic, the testing laboratories, hospitals and quarantine facilities were absolutely outnumbered by the number of new infections arising every day \cite{nacoti2020epicenter}. Moreover, in several countries, a social denial and stigma against testing have been observed \cite{kumar2020covid,torales2020outbreak}  which make identification of patients even more difficult. Thus, it is evident that even with the full healthcare efficiency, promptly detecting and quarantining all the patients is not feasible. In such a condition, the people who have been detected and quarantined, and the people outside quarantine facility, will not only have different contact rates with the susceptible class, they might also have have different recovery rates. While those admitted to hospitals and quarantine facilities will have less opportunity meeting the susceptible, they will be recovering faster from the disease. On the other hand, the infected people who are tested false negative, not yet tested, or awaiting testing results are generally being requested to stay home and self-quarantine themselves. These people might recover slowly and meeting more susceptible people during their recovery period.\\
Beyond these two major divergences, we must also consider the stochastic nature of the real-world dynamics. Starting from exposure to the virus to the detection of the disease, everything is not deterministic, but probabilistic. Moreover, in ODE models, the homogeneous mixing assumption dilutes all spatial information which is essential for modeling a disease which spreads from close-contacts only. While compartmentalised mean-field modeling approaches have their own benefits, to take into account of heterogeneity and spatial infection spread, cellular automata based lattice models provide a powerful tool \cite{toffoli1987cellular,wolfram2018cellular}. Despite the ability to show extremely complex macroscopic outcomes, this modeling tool based on local interactions trusts on the interaction of a multitude of single individuals, giving a direct correspondence to the physical system \cite{chopard1998cellular,sante2010cellular}. In PCA, the transitions can be considered based on certain pre-defined probabilities \cite{mairesse2014around} which also suits the current dynamics of this global pandemic. Thus, PCA based modeling will have rich information about the stochastic spatiotemporal spread of the infection in a population. \\
With these understanding of COVID-19, we proceed to model the spread of this infectious disease using a spatially explicit epidemiological lattice model using PCA. Here the spatial distribution of population is characterized by a set of discrete lattice points. Specific probabilistic rules are made to define how every lattice point will change its state through the possible interactions between the individuals and movement. We use rigorous simulations using PCA to understand the effects of lockdown implementation, regulatory compliance, testing efficiency and availability of healthcare facilities. To the best of our knowledge, this is the first attempt to model COVID-19 using cellular automata keeping spatial population distribution in mind. We exhibit that we are able to produce the temporal information about the infection, along with the visualization of the space that is absent in ODE based modeling. The article is organized as follows. In Section 2, we will define and discuss the PCA model of COVID-19 spread. Next, in Section 3, we will present and analyse the simulation results, considering probabilistic dynamics in each step of disease progression, i.e., exposure, proliferation and quarantining. Finally, in Section 4 major conclusions relating model results with the real epidemic data from different countries and future scopes will be discussed. 

\section{Model Description}
\label{sec:1}
In this section, we define a spatially explicit epidemiological model through PCA on a square lattice with SEIQR structure. We consider that, instead of having three subpopulations, COVID-19 has partitioned our society in five different subpopulations. While $S$ and $R$ have the same implications as before ($S$: susceptible, $R$: recovered and removed), the $I$ subpopulation is divided into three segments: those who are exposed and asymptomatically infected, denoted by $E$; those who have shown symptoms, but are not detected/tested positive/quarantined yet, denoted by $I$; and those who are quarantined or hospitalized, $Q$. \\
Consider a community which has been infected by the virus.  Let us consider a two-dimensional lattice of dimension $L \times L$, where an SEIQR epidemic process is going on. Each small box or \textit{`patch'} in this lattice can be occupied by a person (this however can be generalised into a house or residential area) or can remain empty (black in Fig. \ref{fig:time}). The occupied patches have to belong to any one of the states, \textit{S} (blue), \textit{E} (green), \textit{I} (red), \textit{Q} (cyan) or \textit{R} (yellow). In all the simulations, the neighborhood condition chosen as per the standard rules of Moore neighborhood \cite{davies1995effect}, in general, if not precisely mentioned otherwise.  Let us setup the following probabilistic rules for PCA to establish the logic behind transmission of the disease:
\begin{itemize}
    \item \textbf{Rule 1.} Every person is surrounded by at most eight neighbors. If any one of these neighbors are exposed or infected (belongs to \textit{E} or \textit{I} class), then the person has a finite probability to get exposed
    \item \textbf{Rule 2.} Exposed person is capable of infecting susceptible people.
    \item \textbf{Rule 3.} Exposed people (\textit{E}) have a finite probability to show the symptoms of the disease and become infected (\textit{I}). Infected people can also spread the infection to a susceptible person.
    \item \textbf{Rule 4.} Once a person gets infected, depending on the efficiency of the testing process in a certain period of time and with a certain probability the person gets detected, and thus quarantined, depending upon the available health infrastructures. A quarantined person is restricted to come in contact with susceptible further. 
    \item \textbf{Rule 5.} No birth or immigration is considered, keeping the total number of people in the population constant.
    \item \textbf{Rule 6.} We do not consider death due to the disease separately in our model; death is incorporated inside the recovered class and considered to be removed from the population.
    \item \textbf{Rule 7.} The movement of each of these people is implemented by considering a sphere of influence which establishes a boundary in which the person can meet any other person and come in direct contact, which can finally result in transmission of the disease. A susceptible person cannot get infected or exposed, if his sphere of influence does not contain any exposed or infected person.
\end{itemize}
With this rules setup, the framework can be considered as a combination of  transition probability model \cite{sacks1977transition} and discrete event model \cite{dean2014discrete} where the disease transmission can set in through several subevents:
\begin{itemize}
    \item \textbf{Getting exposed:} $S \longrightarrow E$, with effective rate $p_e [n_E(i)+ n_I(i)]$  where $p_e$ is the exposure probability; this transition rate is also proportional to the total number of exposed and infected neighbors $n_E(i)$ and $n_I(i)$ of the $i^{th}$ susceptible site, as a susceptible person will meet its exposed/infected neighbors independently.
    \item \textbf{Getting infected, with symptoms:} $E \longrightarrow I$, with transition probability $p_i$, which estimates the probability that the infection will grow and show substantial symptoms. A threshold time of $t_i$ is associated with this subevent to account for the incubation period, which is a very prominent problem in case of COVID-19.
     \item \textbf{Getting quarantined:} $I \longrightarrow Q$, takes place after a threshold time of $t_q$ and  with a probability $p_q$ which measures the promptness of testing and efficiency of testing methods respectively.
     \item \textbf{Recovery or removal:} $I \longrightarrow R$ and $Q \longrightarrow R$ takes place after an average time of $t_r$ with a probability $p_r$ which takes into account the effectiveness of health infrastructure and medical facilities.
\end{itemize}
Fig. 1 compiles all these events together in the form of a disease spread model. The total population $N$ is fixed while the simulation is carried out, but its variations estimate the population density of the zone of interest.
\begin{figure}
  \includegraphics[width=\textwidth]{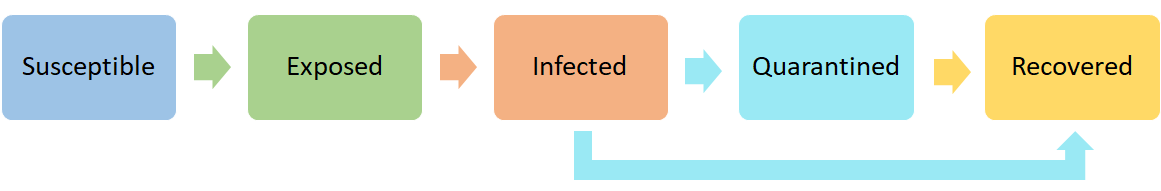}
\caption{The transitions related to the subevents in the epidemic spreading model.}
\label{model}       
\end{figure}
\section{Simulation and Analysis}
\label{sec:2}
We perform simulations on a homogeneous two-dimensional landscape with $400 \times 400$ patches with $100000$ randomly selected patches as residences of people in a society. Fig. 2 (a) shows a zone of interest at initial condition, where blue patches denote susceptible individuals. Considering Moore neighborhood (8 neighborhoods) a synchronous updating for all sites is carried out. However, the neighbourhood conditions are modified accordingly to estimate the allowed interactions for a person. Parameter $d$ estimates the distance up to which the interactions are allowed during the implementation of simulation. The interaction convention (shown in Fig. 2 (c)) has been set as follows: if a person is staying home, then $d = 0$; if the sphere of influence includes only the first nearest neighbours, then $d = 1$; if the sphere of influence includes first as well as next nearest neighbours, $d = 2$, and so on. For better visualization of the dynamics, we normalize any subpopulation with the total number of people in the entire population. We have used lowercase symbols to represent normalized subpopulation, e.g. the normalized fraction of people infected in the entire population is represented with `$i$', whereas the normalized fraction of people quarantined in the population is represented with `$q$' and so on. As incubation period for COVID-19  \cite{lauer2020incubation}, $t_i$ is associated with the infection itself, so we keep it constant for all our experiments with default value of $t_i=8$. The simulations are run until a saturation is reached. 
\begin{figure}
  \includegraphics[width=\textwidth]{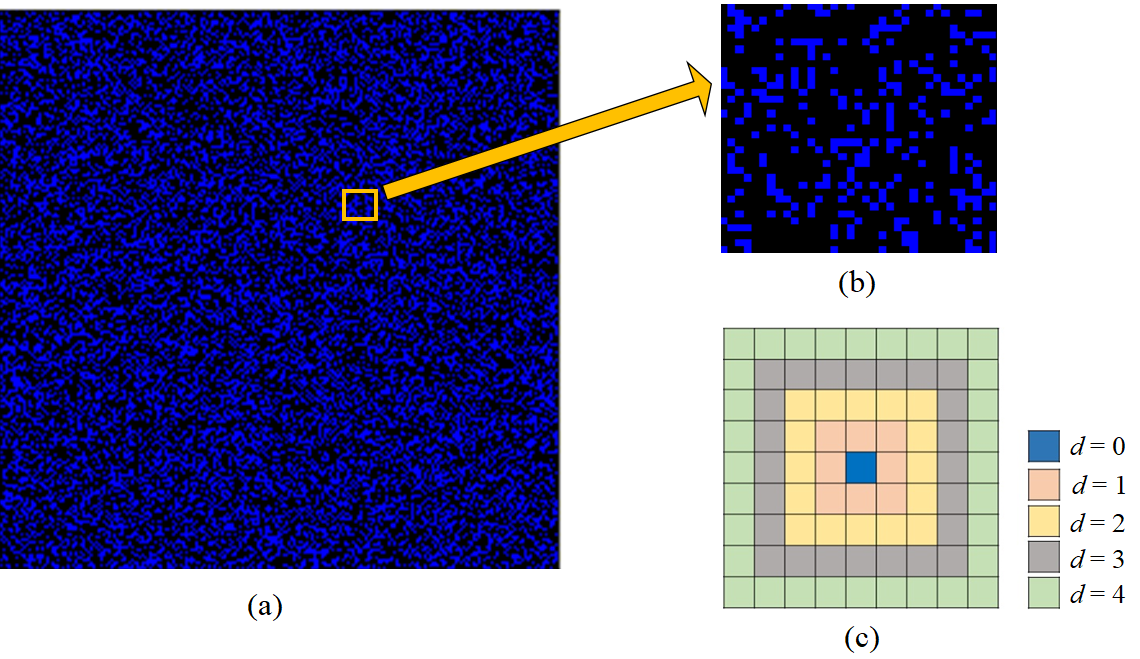}
\caption{Simulation landscape and interaction neighbourhood: (a) A zone of interest with $400 \times 400$ patches. Density of small blue boxes represent population density. (b) A small sub-zone, denoting the distribution of susceptible and empty patches. (c) Interaction conventions: staying home, $d = 0$; sphere of influence includes only first nearest neighbours denoted by $d = 1$; sphere of influence includes first as well as next nearest neighbours denoted by $d = 2$, and so on.}
\label{d}       
\end{figure}
\subsection{Effects of population density and movement restriction}
\label{without}
\begin{figure}
\center
  \includegraphics[width=\textwidth]{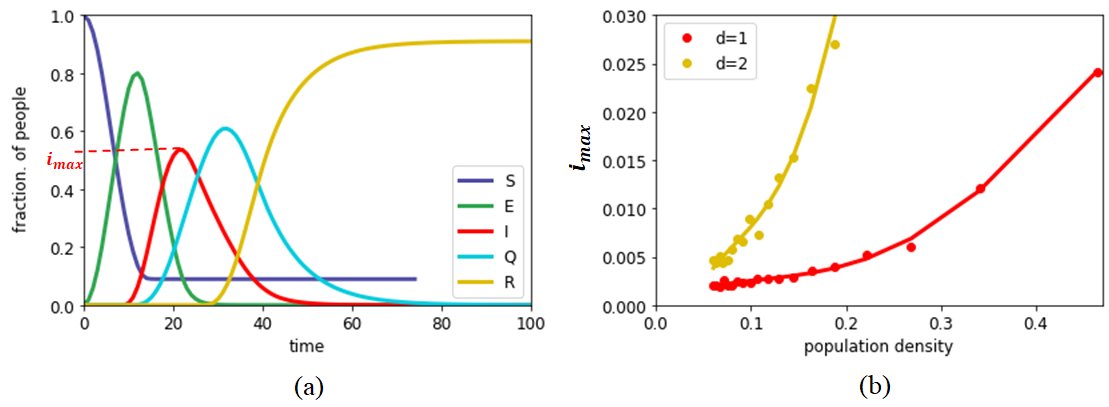}

\caption{Dynamics of the infection spread: (a) time evolution of five subpopulations for $d=2$, $p_{e}=0.5$, $p_i=0.5$, $p_q=0.1$ and $p_r=0.12$ and population density $D=0.46$; (b) Effect of population density in infection spread. Here population density $D=1$ indicates that all the lattice locations are occupied by people. Other parameters are kept same as before. }
\label{fig:density}      
\end{figure}
\begin{figure}
\center
  \includegraphics[width=\textwidth]{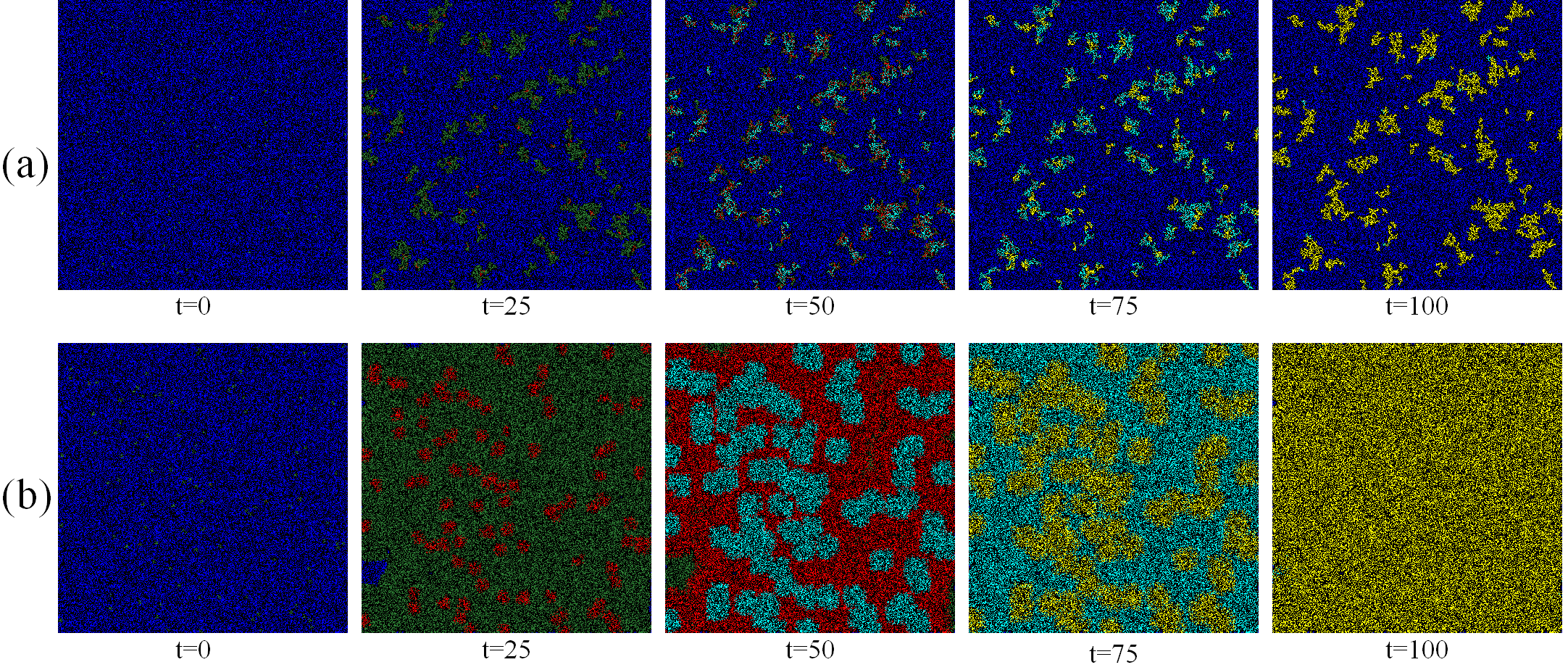}
\caption{Spread of the infection in a population with different parameters as mentioned in Sec. \ref{sec:2} for (a) $0\leq d\leq 1$ and (b) $0\leq d \leq 3$. The colors of the subpopulations are same as shown in Fig. \ref{model}. All the transition probabilities are same as mentioned in Sec. \ref{prompt}.   }
\label{fig:time}       \end{figure}

The density of a particular place may play a key role in the spreading dynamics as it is difficult to implement strict social distancing measures in densely populated areas. It can be observed that cities with large number of residences have often become the hot-spot of infection spread. To understand the effect of density, we simulate our model with different population densities, and examine how the maximum fraction of infected people in the population depends on it. As shown in Fig. \ref{fig:density}, the population density has a severe impact on the fraction of maximum infection ($i_{max}$). As the population density increases, the $i_{max}$ increases exponentially. This happens as the person-to-person transmission probability increases drastically, and even a very restricted movement ($d=1$) helps to spread the infection in the community. It can be observed that for $d=2$, the $i_{max}$ increases much more rapidly as population density increases. Thus, to stop the maximum reach of the infection, it is necessary to administer strict movement control, or lockdown.\\
In our simulations, lockdown imposes a restriction on movement of every person on the lattice.  In Fig. \ref{fig:time}, we have shown time evolution of two different communities to demonstrate the effect of lockdown implementation with the help of parameter $d$. We assume that for a given $d$, a person may meet someone to the $d^{th}$ neighbourhood but comes back at his original place at the end of each step. Thus, the demography does not change in each step and the neighbourhood information remains constant. In Fig. \ref{fig:time}(a), a strict lockdown is imposed ($0\leq d \leq 1$), where staying at home or  occasional movement within $d=1$ neighbourhood in urgency is allowed. Here we observe that, after saturation, most of the people remain in the susceptible state showing a very limited spread of the infection. On the other hand, In Fig. \ref{fig:time}(b), the movement in the community was not restricted, and every individual was allowed to move as per their habit/choice ($0\leq d \leq 3$). In this case, it can be observed that almost all the people come in the contact of the pandemic.  Thus, restricting the movement emerges as a highly impactful strategy for controlling the infection spread.\\
\subsection{Promptness in implementing lockdown:} 
\label{prompt}

As discussed in Sec. \ref{without}, lockdown is necessary to restrict the spread of the infection in a community. However, implementation of a total lockdown is a big challenge for a country because of several socioeconomic issues. An abrupt lockdown may create panic and associated complications as well. However, in our studies, by introducing a delay time parameter $t_L$, we found that an early implementation of lockdown is extremely useful to control the spread of the disease. To observe this, members of the populations were allowed to interact with $0\leq d\leq 2$ initially. If at $t_0$ time the infection first enters the community with $E_{int}$ and $I_{int}$ number of initially exposed and infected people, then we assume that a lockdown gets implemented at $t_0+t_L$ time step.  For this experiment, we considered  $I_{int}=6$, $E_{int}=200$, $p_e=0.5$, $p_i=0.5$, $p_q=0.1$ and $p_r=0.12$. For a probable infected person, quarantining and recovery threshold times are taken as $t_q=2$ and $t_r=18$. As shown in Fig. \ref{fig:fig4}(a), as $t_L$ increases, i.e., as the delay in implementing the lockdown increases, the fraction of people getting infected increases drastically. Thus, even with the difficulties of immediate implementation, it is necessary to plan for an early lockdown to restrict movements in case of a pandemic.
\subsection{Effect of initial infection level:}
\begin{figure}
\center
  \includegraphics[width=\textwidth]{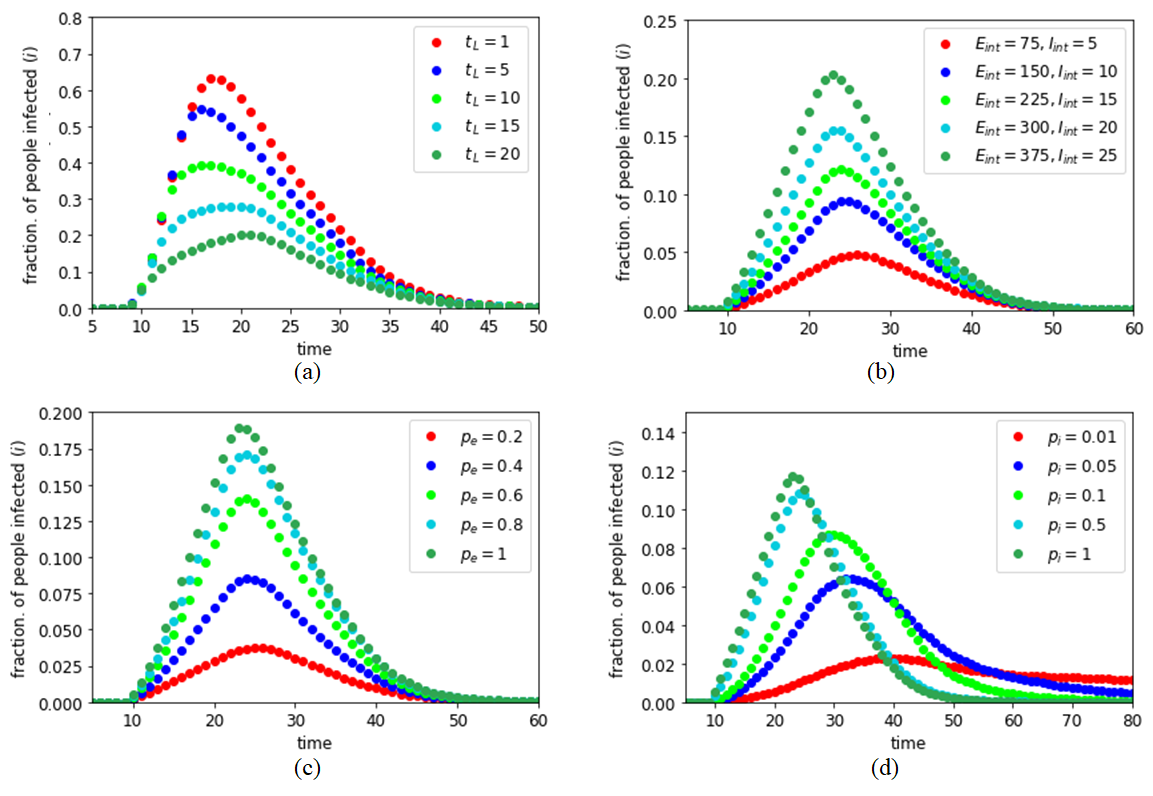}
\caption{Dynamics of the infection spread in lockdown: (a) Effect of delay in implementing lockdown; (b) Effect of different states (initial conditions) just before lockdown is implemented; (c) Effect of exposure probability $p_e$;  (d) Effect of infection probability $p_i$. }
\label{fig:fig4}       \end{figure}
The number and location of the infected and exposed people in the population just before the start of the lockdown greatly influence the maximum spread of the infection in the community. As shown in Fig. \ref{fig:fig4}(b), as $E_{int}$ and $I_{int}$ increase, the maximum spread of infection increases keeping the duration of the infection almost constant. For  Fig. \ref{fig:fig4}(b), we considered that $t_L=0$, i.e., the strict lockdown is enforced immediately in the population with $p_e$, $p_i$, $p_q$, $p_r$, $t_q$ and $t_r$ as mentioned in Sec. \ref{prompt}. This experiment depicts that it is strongly needed to control the initial infection to make the lockdown effective. Any act, e.g. migrant movement, social gathering, unhygienic practices in public places etc.,  should be handled carefully to restrict the initial infection level until the lockdown can be implemented in the society.
\subsection{Effect of exposure probability:} An effective way to fight a pandemic like COVID-19 is to reduce the exposure probability $p_e$  as reduction of the exposure probability reduces the chance of getting infected at the first place. Our experiment also depicts that $p_e$ is an important factor in controlling the spread. As shown in Fig. \ref{fig:fig4}(c), when $p_e$ reduces from 0.8 to 0.2, the maximum spread of infection $i_{max}$ reduces almost linearly. In this experiment, we considered  all the parameters, except $p_e$, same as mentioned in Sec. \ref{prompt} with $t_L=0$. It shows that while an infected person may meet a healthy person even in a lockdown condition, social distancing and other protective measures like usage of facemask, sanitizer etc. reduce the possibility of contamination. Thus, social distancing and other precautions may act as social vaccine to prevent the disease spread.
\subsection{Effect of infection probability:} Depending on the nature of the infectious disease and the immunity strength of the members of a community, infection probability $p_i$ plays a major role in the infection spreading. In a pandemic scenario, $p_i$ may vary for various reasons like mutation of the virus, acquired immunity or simply because of the low infection probability of the disease. As shown in Fig. \ref{fig:fig4}(d), as $p_i$ reduces, $i_{max}$ reduces significantly. However, the reduction of $i_{max}$ is not linear with $p_i$. To understand the effect of $p_i$, we define another parameter called the `infection lifetime ($\tau$)' which is defined by the time difference between the instances when the first person gets infected in the population  and the last person recovers from the infection. Fig. \ref{fig:fig4}(d) thus suggests that $\tau$ increases as $p_i$ decreases. This happens as a person once exposed to the infection remains in the  exposed state. A lower value of $p_i$ reduces someone's chances to get quarantined and also indicates possibility of a delayed symptomatic transition from \textit{E} to \textit{I} class which increases $\tau$. In Fig. \ref{fig:fig4}(d), we considered all the parameters, except $p_i$, same as mentioned in Sec. \ref{prompt} with $t_L=0$.
\begin{figure}
\center
  \includegraphics[width=\textwidth]{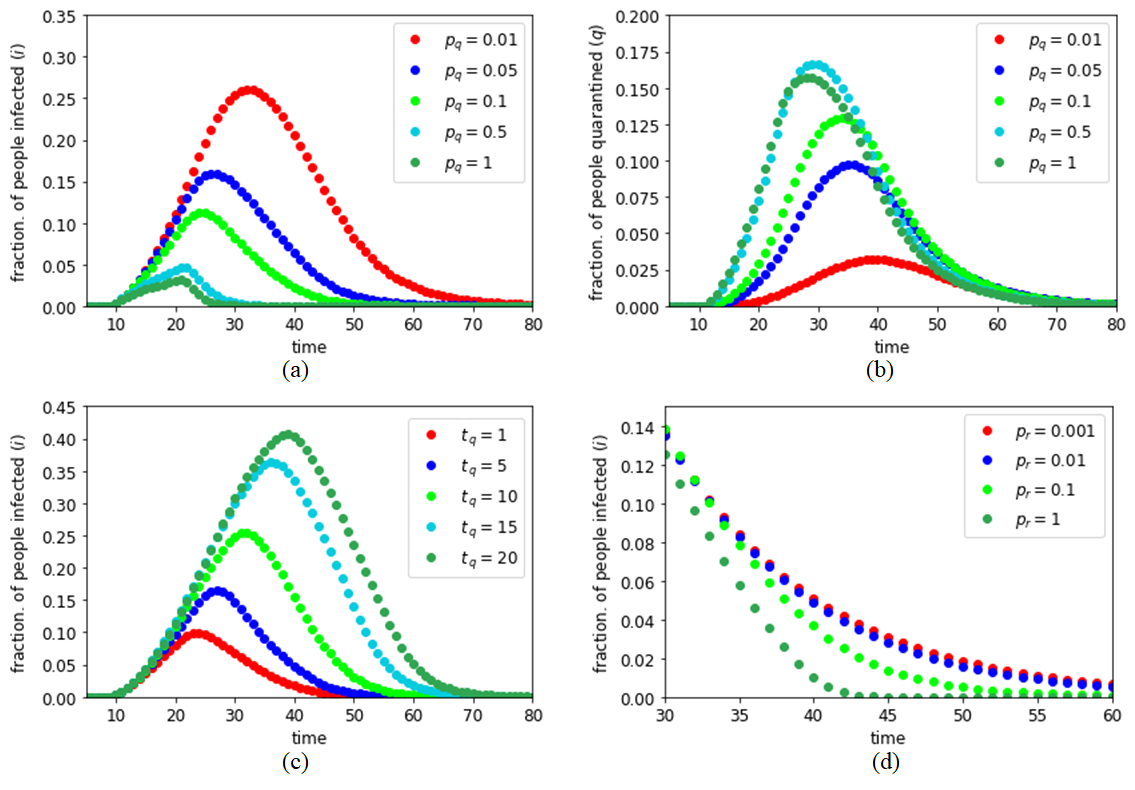}
\caption{Dynamics of the infection spread in lockdown: Effect of testing probability $p_q$ on (a) the fraction of people infected and (b) the fraction of people quarantined; (c) Effect of delay in individual quarantine process; (d) Effect of recovery probability $p_r$. }
\label{fig:fig5}     \end{figure}
\subsection{Effect of testing efficiency:}
In a pandemic like COVID-19, SARS etc., detecting the infected individuals plays a crucial role in controlling the infection spread. Detection of infected people has two major factors$-$ governmental initiatives, and the efficiency as well as supply of the test kits. Moreover, awareness and cooperation of the community also plays a significant role; though, often aware and cooperative infected people are detected in a population, but it is always challenging to detect the infected individuals with denial and resistance, which increases the risk of an outbreak. These three factors are combined in the probability term $p_q$.  We note that this testing efficiency parameter plays a major role in the spreading dynamics. As shown in Fig. \ref{fig:fig5}(a), as $p_q$ increases both $i_{max}$ and $\tau$ decreases which helps to limit the infection.\\
It is important to mention that it is the fraction of quarantined population \textit{`q'} which we can observe as the active cases in a pandemic scenario as the other subpopulations ($e$ and $i$) are yet undetected, thus not directly observable. As shown in Fig. \ref{fig:fig5}(b), the fraction of people quarantined remains quite less if $p_q$ is small. This signifies that less testing efficiency can be mistaken as a smaller spread of the infection. Thus, to get a clear picture about the spread of the infection in a community, it is recommended to keep $p_q$ high, which in turn also helps to limit the infection spread. In both these figures (\ref{fig:fig5}(a) and (b)), we considered all the parameters, except $p_q$, same as mentioned in Sec. \ref{prompt} with $t_L=0$.  
\subsection{Effect of testing time $t_q$ and recovery probability $p_r$:} Testing time also plays a decisive role in epidemic dynamics. The testing time might be dependent on the available facilities and cooperation of individuals to report to the medical centers as soon as they get any symptom of the infection. As shown in Fig. \ref{fig:fig5}(c), as $t_q$ increases, both $i_{max}$ and $\tau$ increase as the infected people get more scope to spread the infection in the community. Thus, early reporting of any viral symptom and test kits that can produce test results fast are essential to limit a pandemic. In Fig. \ref{fig:fig5}(c), we considered all the parameters, except $t_q$, same as mentioned in Sec. \ref{prompt} with $t_L=0$.\\
Though recovery probability largely depends on the disease itself, but depending on the medical support and other available health services, recovery probability might vary in a population. We found that recovery probability $p_r$ does not have much impact on $i_{max}$ as the spread of an infection is usually faster than the recovery time, but, as shown in Fig. \ref{fig:fig5}(d), it limits the infection lifetime $\tau$. Thus, good health facilities and medical supports are required throughout the pandemic period to speed up the recovery process if possible. In Fig. \ref{fig:fig5}(d), we considered all the parameters, except $p_{r}$, same as mentioned in Sec. \ref{prompt} with $t_L=0$.
\section{Discussions}
\label{sec:4}
In this prevalence of pandemic COVID-19, ODE based kinetic models are being trusted extensively for the purpose of predicting time dependent profiles and steady state behaviour of the epidemic dynamics. In this paper, we argue that cellular automata based simulations can provide a more realistic and insightful platform for modeling this dynamics. It is undeniable that a substantial variability has been observed in the nature of the course of the disease in countries and provinces all around the world. While a fully operational clinically trialed vaccine can only be expected after a year or so for everyone, several countries are struggling hard to find a proper mitigation strategy for the disease. It is interesting to see that publicly available data of confirmed daily cases for different countries are quite different from each other. While in some countries the disease has completed its course in around 60 days (like, Germany, Switzerland, Singapore etc.), in some other countries it is taking more than 80 days to reach the peak (like, Italy, India, Brazil etc.). In countries like Spain, Portugal and Turkey, a distinct departure from symmetric bell-shaped behaviour of active cases can be observed, which is heading towards a saturation, indicating a substantial persistence of the disease. For Iran, emergence of a second peak is imminent, which cannot be accounted for using ordinary differential equation based SIR models. Many countries have shown equivalent rise of the epidemic in initial stages are showing drastic divergence in the later stages. The applicability of strategies like aggressive testing, which found their success in countries like  Singapore, Hong Kong and South Korea, are doubted for the high population countries like, India, USA etc. \\
To analyse and interpret this data, it must be considered that along with the parameters associated with infection itself, the dynamics reflects the human interventions and reactions as well, which are highly probabilistic. In this paper, we have used a lattice-based PCA epidemiological modeling to efficiently mimic a pandemic environment and find out the contributions of each major parameter associated with the pandemic spread. We identified the role of different factors like probability of exposure, testing probability, promptness of implementing lockdown etc. which can not only be used to design better strategies to control a pandemic like COVID-19, but also can be used as analyzing tools to understand and explain proposed initiatives like social distancing, usage of facial mask, importance of sanitizers etc. Our exhaustive analysis shows that countries with high population density should handle lockdown and migration carefully as they have higher risk of getting infected. Using quantities like $i_{max}$ and $\tau$, we have also characterised and quantified the independence of rise and fall of the distribution, showing the clear influence of human intervention strategies. Our model individually associates tunable parameters with several practical factors like population density, testing efficiency, public awareness, health-care facility, general immunity etc. which are implicit to a country, and can contribute significantly in the country-wise data variations. By demonstrating the flexibility of the model, we conclude that this is not only capable of explaining the reasons behind divergence of epidemic dynamics in different countries, but also has the potential to explain unforeseen nature of a new data that might depend strongly on space information, behavioural probabilities or underlying fluctuations. Our study establishes a prescription about how the resultant distribution changes its properties (e.g., peak position, sharpness of rise, lifetime of epidemic, asymmetric long-tailed fall etc.) driven by each of these parameters. As this simulation-based study explicitly includes the spatial factors into the dynamics, incorporating realistic behavioural and demographic features is achievable through this model. Thus, our methodology proposes a more accurate and flexible platform to understand the country-wise diversity of the observed data than the mean-field models.
 \section*{Conflict of interest}
The authors declare that they have no conflict of interest.

\bibliographystyle{spmpsci}      
\bibliography{bibliography}   
\end{document}